\documentstyle[12pt,twoside]{article}
\def\Rightheadtext{Consistent Interactions Between Gauge Fields}
\def\Leftheadtext{Marc Henneaux}

\def\keywords#1{\footnotetext{{\it Key words and phrases:\/} #1}}
\def\classification#1{\footnotetext{#1}}
\def\Abstract#1{{\leftskip1cm\rightskip1cm\footnotesize\noindent
{\sc Abstract.} #1 \par}}
\font\bbf=cmbxti10 scaled\magstep1
\makeatletter
\def \section{\removelastskip\@startsection{section}{1}
  {\z@}
  {-3.5ex plus-1ex minus-.2ex}
  {2.3ex plus.2ex}
  {\def\@svsec{\thesection.\space}\noindent\hfil\sc}}
\def \subsection{\removelastskip\@startsection{subsection}{1}
  {\z@}
  {-3.5ex plus-1ex minus-.2ex}
  {2.3ex plus.2ex}
  {\def\@svsec{\thesubsection.\space}\noindent\hfil\sc}}
\makeatother

\def \thebibliography#1{\section*{References}
  \frenchspacing
  \raggedbottom
  \footnotesize
  \list{[\arabic{enumi}]}{\settowidth\labelwidth{[#1]}
    \leftmargin\labelwidth \advance\leftmargin\labelsep
    \itemsep 0pt \parsep 0pt
    \usecounter{enumi}}
  \def \newblock{\hskip .1em plus .3em minus .07em}
  \sloppy\clubpenalty10000\widowpenalty10000}

\begin{document}
\setcounter{page}{1}

\def\Refs{\list{}{\topsep-20pt\itemsep0pt\parsep0pt}\item[]}
\let\endRefs=\endlist
\def\References{\begin{center}
{\sc References}
\end{center}
\vspace*{-17.5pt}
\footnotesize}
\newcount\firstpage
\firstpage\thepage
\makeatletter
\renewcommand{\@oddhead}
   {\ifnum\thepage=\firstpage{}\else{\hfil\sc
    \lowercase\expandafter{\Rightheadtext} \hfil\thepage}\fi}
\renewcommand{\@evenhead}
   {\ifnum\thepage=\firstpage{}\else{\thepage\hfil\sc
\lowercase\expandafter{\Leftheadtext} \hfil}\fi}
\long\def\@makefntext#1{\parindent 1em\noindent
                   \ifnum\@thefnmark=0 \hbox to1em{\hss}
                   \else\hbox to1.8em{\hss$\m@th^{\@thefnmark}$}\fi #1\message{footmark=\@thefnmark}}
\makeatother

\begin{flushright}
\bf International Conference on\\
\bbf Secondary calculus and cohomological Physics,\\
\bf Moscow, August 1997
\end{flushright}

\classification{ }


\vskip.5cm

\begin{center}
\uppercase{\bf Consistent Interactions Between Gauge Fields:
The Cohomological Approach}
\\ \vspace{1.084cm}
\sc Marc Henneaux\\  \vspace{.373cm}
\rm Universit\'e Libre de Bruxelles\\ Campus Plaine C.P. 231, B-1050
Bruxelles, Belgium\\
{\it E-mail:\/} henneaux@ulb.ac.be \\
\end{center}
\vskip.576cm

\Abstract{The cohomological approach to the problem of
consistent interactions between fields with a gauge freedom
is reviewed.  The role played by the BRST symmetry is explained.
Applications to massless vector fields and $2$-form gauge
fields are surveyed.}

\section{Introduction}
\setcounter{equation}{0}

The BRST symmetry was originally discovered in a purely quantum
context \cite{BRS,Tyutin}.  It was realized only later
that it has also
a useful classical interpretation.  This was done first
in the Hamiltonian context developed by
Fradkin and his school \cite{FV,BV1,FF,BF1}, where it was
shown that the BRST cohomology
can be related to the Hamiltonian reduction
of the system \cite{H2,HT1,DV,McMull,Brow,FHST,Sta}.
This cohomological understanding of the BRST symmetry
enabled one to provide as a by-product complete proofs
of the existence of the BRST generator for an
arbitrary gauge system (subject to definite regularity
conditions) \cite{FHST,HT}, paying due account to
the global phase space features (the original proofs
\cite{H2,BV2,BF2},
based on some particular local representation of the
constraint surface, were only local).

More recently, it has been observed in the Lagrangian context
that the classical problem
of consistently introducing interactions in a gauge
theory can also be usefully reformulated in terms of
the BRST differential and the BRST cohomology \cite{BH}
(see also \cite{Stasheff}).  The use of cohomological ideas systematizes 
the search for all possible consistent interactions and, moreover,
relates obstructions to deforming a gauge-invariant action
to precise cohomological classes of the BRST differential.

The purpose of this review article is to survey the BRST
approach to the problem of consistent interactions.
Applications of the general theory to massless vector fields
and $2$-form gauge fields are also reviewed.  Other
applications (with references to
the original literature) are listed at the end.  

\section{The problem of consistent interactions between gauge fields}
\setcounter{equation}{0}

\subsection{Consistent interactions for a set of $U(1)$-gauge fields}

We shall develop the  theory mostly by means of examples.
Consider a set of $N$ massless abelian vector fields $A^a_\mu$ described
each by the familiar Maxwell action.  The free 
action is thus
\begin{equation}
I_0 = -\frac{1}{4} \int d^n x F^a_{\mu \nu} F^{\mu \nu}_a,
\; \; a= 1, \dots ,N
\label{free1}
\end{equation}
with
\begin{equation}
F^a_{\mu \nu} = \partial_\mu A^a_\nu - \partial_\nu A^a_\mu.
\end{equation}
The action (\ref{free1}) is invariant under the gauge
transformations
\begin{equation}
\delta_\epsilon A^a_\mu = \partial_\mu \epsilon^a,
\label{freegauge1}
\end{equation}
which close according to an abelian algebra.
The gauge symmetry of (\ref{free1}) is quite important since
it removes the unphysical (longitudinal and temporal)
degrees of freedom. The free equations of motion are
\begin{equation}
\frac{\delta I_0}{\delta A^a_\mu} = \partial_\nu F_a^{\nu \mu}
= 0.
\label{freeEOM1}
\end{equation}
We shall assume that the spacetime dimension is strictly greater
than $2$ (in $2$ dimensions, the theory has no local
degree of freedom).

The question addressed here is whether one can
add interaction terms to the action (\ref{free1}) in
a manner that maintains the number, but not necessarily the form,
of the gauge transformations.  In other words, we want to deform
the free action by adding to it interaction terms
\begin{equation}
I_0 \rightarrow I_0 + g I_1 + g^2 I_2 + \dots
\label{deformation1}
\end{equation}
and to deform simultaneously the gauge symmetries
\begin{equation}
\delta_\epsilon A^a_\mu = \partial_\mu \epsilon^a
+ g  \Delta^a_{\mu b} \epsilon^b + O(g^2)
\label{gaugedeformation1}
\end{equation}
in such a way that the deformed action is invariant under
the deformed gauge transformations, at each order in the
``coupling constant" $g$.  The expansions (\ref{deformation1})
and (\ref{gaugedeformation1}) are a priori
formal power series in $g$
and we shall not worry about convergence questions here.  In most
cases, however, the series terminate or can be made to terminate upon
introduction of appropriate auxiliary variables.
It is required that each term in the expansion (\ref{deformation1})
be a local functional, i.e., be the integral of a function
of the vector potentials and their derivatives up to some
finite order.  

We insist that the deformed action should have the same number of
gauge symmetries as the original, free, action, because 
the decoupling of the temporal and longitudinal modes,
guaranteed by
gauge invariance, appears to be essential for consistency.  
Although there is
no theorem stating that this is the only possibility leading to a 
consistent theory, we shall consider only this case here.  Other
additional criteria may be imposed on the deformation (e.g. causal
propagation, or preservation of some specific rigid 
symmetry\footnote{The inclusion of rigid symmetries can actually
be also performed along BRST lines, see \cite{BBW}.}) but we shall
focus here only on gauge invariance.  We shall call
throughout ``consistent deformations" the
deformations that preserve gauge invariance
(possibly in a deformed way).

Consistent interactions are easily constructed by taking
for the interaction terms $I_i$ ($i > 0$) functions of the
curvatures $F^a_{\mu \nu}$ and their derivatives.  An example
is given by Born-Infeld theory \cite{BornI}.  Because
these terms are gauge-invariant under the gauge transformations
(\ref{freegauge1}) of the undeformed theory, they do not
deform the gauge symmetry: the full action (\ref{deformation1})
is invariant under the original gauge symmetries.

More interesting deformations (from the algebraic point of
view) are those that deform
not only the action, but also the gauge transformations
and their algebra.  A well-known example is the
Yang-Mills gauge theory in which the abelian symmetry
(\ref{freegauge1}) is replaced by a non-abelian one.
We shall see below to what extent the Yang-Mills construction
is unique.

\subsection{Consistent Interactions For A Set Of Free Exterior $2$-Forms}

The same problem can be addressed for any gauge system.  In particular, one
may start with the free action describing a system
of exterior $2$-forms $B^A_{\mu \nu}$ instead of (\ref{free1}) 
\begin{equation}
I_0 = - \frac{1}{2 \cdot 3!} \int d^nx H^A_{\mu \nu \rho}
H_A^{\mu \nu \rho}, \; \; A = 1, \dots, N
\label{free2}
\end{equation}
with 
\begin{equation}
H^A_{\mu \nu \rho} = \partial_\mu B^A_{\nu \rho}
+ \partial_\nu B^A_{\rho \mu} + \partial_{\rho} B^A_{\mu \nu}.
\end{equation}
The gauge transformations are now
\begin{equation}
\delta_\epsilon B^A_{\mu \nu} = \partial_\mu \epsilon^A_\nu
- \partial_\nu \epsilon^A_\mu
\label{freegauge2}
\end{equation}
and the equations of motion read
\begin{equation}
\frac{\delta I_0}{\delta B^A_{\mu \nu}} = \frac{1}{2} \partial_\lambda
H_A^{\lambda \mu \nu} = 0.
\label{freeEOM2}
\end{equation}
The new feature of this model, however, is that the gauge transformations
cease to be independent.  Indeed, if one takes as gauge
parameters pure gradients, 
\begin{equation}
\epsilon^A_\mu = \partial_\mu \lambda^A
\label{reducibility1}
\end{equation}
one gets gauge variations of the fields that are identically zero.

Since the deformed theory should have the same number of independent
gauge symmetries as the original one, we require that the
deformation
should also preserve the number of reducibility identities.  That is,
the deformed gauge symmetries should reduce to zero (at least
on shell) when the gauge parameters are given by some definite
deformation of (\ref{reducibility1}),
\begin{equation}
\epsilon^A_\mu = \partial_\mu \lambda^A
+ g \Sigma^A_{\nu B} \lambda^B + O(g^2)
\end{equation}

For the $2$-form model, we shall assume 
that the spacetime dimension is strictly greater
than $3$ (in $3$ dimensions, the theory has no local
degree of freedom).

Again, one may easily construct consistent interactions for the system by
adding terms that involve  only the curvature components
$H^A_{\lambda \mu \nu}$ and their derivatives.  These
interactions do not modify the gauge symmetries (\ref{freegauge2})
since they are strictly gauge invariant.  It turns out that contrary
to the vector case, these are actually the only consistent
interactions in spacetime dimensions $\geq 5$ (with, possibly, Chern-Simons
terms that do not modify either the gauge symmetries, see below).

\subsection{General Equations} 

To write down the general equations and to convey
succintly the main qualitative ideas, 
it is convenient to adopt condensed 
notations.  The fields will be collectively denoted by
$\phi^i$.  So, in the vector case, $\phi^i \equiv A^a_\mu$,
while in the $2$-form case, $\phi^i \equiv B^A_{\mu \nu}$.
The undeformed action  $I_0[\phi^i]$ is a local functional of
the fields, and the Euler-Lagrange equations are
\begin{equation}
\frac{\delta I_0}{\delta \phi^i} = 0.
\label{FreeEOM}
\end{equation}
The gauge symmetries of the undeformed theory are given by
\begin{equation}
\delta_\epsilon \phi^i = R^{(0)i}_\alpha \epsilon^\alpha
\label{Freegauge}
\end{equation}
where there is an implicit integration over spacetime in
(\ref{Freegauge}) -- besides the summation over the
index $\alpha$ --, and where $R^{(0)i}_\alpha$ is linear in
$\delta(x,y)$ and (a finite number of) 
its derivatives (locality of the
gauge transformations).  Thus, (\ref{Freegauge})
really stands for
\begin{equation}
\delta_\epsilon \phi^i(x) = \int d^n y R^{(0)i}_\alpha(x,y) 
\epsilon^\alpha(y)
\label{FreegaugeBis}
\end{equation}
with
\begin{equation}
R^{(0)i}_\alpha(x,y) = r^{(0)i}_\alpha \delta(x-y) + r^{(0)i\mu}_\alpha
\delta,_\mu(x-y) + \cdots + r^{(0)i\mu_1 \dots \mu_s}_\alpha
\delta,_{\mu_1 \dots \mu_s}(x-y).
\end{equation}
The coefficients $r^{(0)i}_\alpha$, ... $r^{(0)i\mu_1 \dots \mu_s}$ are
local functions.  The invariance of the action under the gauge transformations
(\ref{Freegauge}) is equivalent to the so-called
Noether identities
\begin{equation}
\frac{\delta I_0}{\delta \phi^i} R^{(0)i}_\alpha = 0
\label{FreeNoether}
\end{equation}
(identically), where again there is an implicit integration over spacetime
($\int dx \delta I_0 /\delta \phi^i(x)$ $ R^{(0)i}_\alpha(x,y) = 0$).

The deformations of the action and the gauge symmetries are
given by
\begin{eqnarray}
I_0 &\rightarrow& I \equiv I_0 + g I_1 + g^2 I_2 + \dots , \\
R^{(0)i}_\alpha &\rightarrow& R^{i}_\alpha \equiv
R^{(0)i}_\alpha + g R^{(1)i}_\alpha + g^2 R^{(2)i}_\alpha  + \dots. 
\end{eqnarray}
The same locality assumptions are made for the interacting
model, namely, each term in the expansion of the action
is a local functional, and each term $R^{(k)i}_\alpha$ in the expansion
of the gauge symmetries is 
linear in $\delta(x,y)$ and (a finite number of) its derivatives, with
coefficients that are local functions.

Consistency of the deformations holds if and only if the
Noether identities are fulfilled at each order in the deformation
parameter $g$
\begin{equation}
\frac{\delta I}{\delta \phi^i} R^{i}_\alpha = 0.
\label{FullNoether}
\end{equation}
By expanding this condition in powers of $g$, one gets an infinite
number of equations on the deformations of the action and the
gauge symmetries.

If the original theory is reducible, i.e., if there exist
choices of the gauge parameters $\epsilon^\alpha$, say
$\epsilon^\alpha = Z^{(0) \alpha}_A \lambda^A$, such
that the gauge variations of the fields are on-shell trivial,
\begin{equation}
Z^{(0) \alpha}_A R^{(0)i}_\alpha = \mu^{(0)ij}_A
\frac{\delta I_0}{\delta \phi^i}, \; \; \mu^{(0)ij}_A =
- \mu^{(0)ji}_A,
\label{FreeReducibility}
\end{equation}
then, the deformation should preserve this reducibility.  Accordingly,
one should be able to deform both $Z^{(0) \alpha}_A $
and $\mu^{(0)ij}_A$,
\begin{eqnarray}
Z^{(0) \alpha}_A &\rightarrow& Z^{\alpha}_A \equiv 
Z^{(0) \alpha}_A + g Z^{(1) \alpha}_A + g^2 Z^{(2) \alpha}_A + \dots , \\
\mu^{(0)ij}_A &\rightarrow& \mu^{ij}_A \equiv
\mu^{(0)ij}_A + g \mu^{(1)ij}_A + g^2 \mu^{(2)ij}_A + \dots
\end{eqnarray}
in such a way that reducibility identities
of the form (\ref{FreeReducibility}) hold for the full
theory,
\begin{equation}
Z^{ \alpha}_A R^{i}_\alpha = \mu^{ij}_A
\frac{\delta I}{\delta \phi^i}.
\label{FullReducibility} 
\end{equation}
Again, by expanding this condition in powers of $g$, one gets an infinite
number of equations on the deformations of the ``structure functions"
$Z^{ \alpha}_A$ and $\mu^{ij}_A$.

\subsection{Algebra Of The Gauge Transformations}

In the course of the deformation, the algebra of the gauge
transformations may of course get also deformed.  The new
gauge transformations may only close on-shell, even if the
original transformations formed a true algebra.  So, one
has
\begin{equation}
R^j_\alpha \frac{\delta R^i_\beta}{\delta \phi^j}
- R^j_\beta \frac{\delta R^i_\alpha}{\delta \phi^j} = 
C^\gamma_{\alpha \beta} R^i_\gamma + M^{ij}_{\alpha \beta}
\frac{\delta I}{\delta \phi^i}, \; \; M^{ij}_{\alpha \beta} = - 
M^{ji}_{\alpha \beta}
\label{Fullalgebra}
\end{equation}
with
\begin{eqnarray}
C^\gamma_{\alpha \beta} &=& C^{(0)\gamma}_{\alpha \beta}
+g C^{(1)\gamma}_{\alpha \beta} + O(g^2) \\ 
M^{ij}_{\alpha \beta} &=& M^{(0)ij}_{\alpha \beta} + g
M^{(1)ij}_{\alpha \beta} + O(g^2).
\label{Deformedstructure}
\end{eqnarray}
The $C$'s and $M$'s can depend on the fields. 
The structure relations (\ref{Fullalgebra}) are
actually consequences of the Noether identities and of the
fact that the gauge transformations are assumed to form
a complete set (see e.g. \cite{HT}).

In the abelian models considered above, the structure functions
$ C^{(0)\gamma}_{\alpha \beta}$, $M^{(0)ij}_{\alpha \beta}$
(and $\mu^{(0)ij}_A$ in (\ref{FreeReducibility})) all vanish,
but they may no longer vanish in the deformed theory.  We thus
allow a priori for the most general deformation compatible
with the existence of gauge symmetries and do not
impose any restriction on the deformed structure
functions.

\section{Trivial Deformations}
\setcounter{equation}{0}

The above equations on the deformations 
always admit solutions of a
particular type.  These solutions are simply obtained by  
making $g$-dependent
redefinitions of the field variables,
\begin{equation}
\phi^i \rightarrow \phi^i + g \; k^i + 0(g^2)
\label{redefinition}
\end{equation}
where $k^i$ can depend on the fields and their derivatives.
Under such a redefinition, the action and the gauge transformations
are in general modified.  For instance, the action becomes
\begin{equation}
I_0[\phi^i] \rightarrow I[\phi^i] = I_0[\phi^i + g \; k^i + 0(g^2)] =
I_0[\phi^i] + g \frac{\delta I_0}{\delta \phi^i} k^i + O(g^2)
\label{trivialdeformation}
\end{equation}
and a similar redefinition holds for the gauge transformations.
Such transformations are, however, rather trivial, and will be
regarded in the sequel as ``fake" deformations.  Our interest
lies in the determination of the non-trivial 
deformations of 
the action, i.e., in the deformations that do not arise from
a (local) redefinition of the field variables.

It should be noted that even if the fields are
unchanged, there is some freedom in the description of
the gauge transformations.  Indeed, one may always redefine
the $R^{(0) i}_{\alpha}$ as
\begin{equation}
R^{(0) i}_{\alpha} \rightarrow \epsilon^\beta_\alpha R^{(0) i}_{\beta}
+ t^{ij}_\alpha \frac{\delta I_0}{\delta \phi^j}
\end{equation}
where $\epsilon^\beta_\alpha$ and $t^{ij}_\alpha$ are
local functions with $\det \epsilon^\beta_\alpha \not= 0$ and
$t^{ij}_\alpha = - t^{ji}_\alpha$.  Expanding $\epsilon^\beta_\alpha$
and $t^{ij}_\alpha$ in powers of $g$ yields
\begin{eqnarray}
\epsilon^\beta_\alpha &=& \delta^\beta_\alpha + g 
\; \epsilon^{(1)\beta}_\alpha
+ O(g^2), \\
t^{ij}_\alpha &=& 0 + g \; t^{(1)ij}_\alpha + O(g^2), \\
R^{(0) i}_{\alpha} &\rightarrow& R^{(0) i}_{\alpha} + g \big(
\epsilon^{(1)\beta}_\alpha R^{(0) i}_{\beta} + 
t^{(1)ij}_\alpha \frac{\delta I_0}{\delta \phi^j} \big) + O(g^2).
\end{eqnarray}
Because of this ambiguity, the deformation of the gauge
symmetries is not unique (for a fixed deformation of the
action).  

It is of particular interest
to examine the deformations of the action that truly deform
the gauge transformations, i.e., such that there is no
redefinition of the field variables and of the $R$'s
that brings the deformed gauge functions $R^i_{\alpha}$
to the original form $R^{(0) i}_{\alpha}$.  Among the interactions
that deform non trivially the gauge transformations, 
it is customary to distinguish between those that do not
deform the gauge algebra ($C^\gamma_{\alpha \beta}$ 
unchanged), and those that 
(non trivially) do ($C^\gamma_{\alpha \beta} \not= 
C^{0\gamma}_{\alpha \beta}$).

Finally, we note that there is also some ambiguity
in the reducibility functions (when there is reducibility)
which are defined, for a fixed set of field variables
and gauge functions $R^i_\alpha$, up to
\begin{equation}
Z^\alpha_A \rightarrow t^B_A Z^\alpha_B + k^{\alpha j}_A
\frac{\delta I_0} {\delta \phi^j}.
\end{equation}
A non trivial deformation of the reducibility functions is one that
cannot be brought back to the original form by means 
of the allowed redefinitions.

\section{Analysis Of The Equations - Overview}
\setcounter{equation}{0}

The theoretical problem of determining consistent deformations
of a given gauge invariant action is of course not a new
one and has been much studied in the context of consistent
interactions for massless, higher spin, particles.  It
has been formulated in general terms in \cite{Berends1,Berends2}.
The usefulness of the deformation point of view
(but not in the general framework
of the antifield formalism, which allows off-shell open deformations
of the algebra) has been advocated in \cite{Julia}.

The equations determining the consistent deformations are rather
intricate because they are non linear and involve simultaneously not
only the deformed action, but also all the deformed structure constants.
The problem is further complicated by the fact that one
has to sort out the trivial deformations from the non
eliminable ones.

In practice, one first determines the consistent first-order
deformations $I_1$,
which may or may not exist. If they
do, the crucial test is then to determine whether the deformation
can proceed to the next order.  

The cohomological approach systematises the recursive
construction and relates the  consistent interactions
to cocycles of the BRST differential.  Furthermore, trivial
deformations are also trivial in the cohomological
sense, i.e., BRST-coboundaries.
Thus, the two aspects involved in the construction
of consistent interactions (consistency conditions and
elimination of trivial solutions) have a clear cohomological
counterpart and are just the two familiar aspects involved in
computing cohomology (impose coboundary condition
and eliminate trivial solutions, i.e., coboundaries).

Of course, the cohomological approach ultimately deals
with the same equations as the standard approach,
but it organizes them in a rather
appealing way.  Furthermore, it clearly exhibits
the obstructions to deforming the given action to non-trivial
BRST-cohomological classes.  Finally, by reformulating the
problem of consistent interactions as a cohomological
problem, one can bring in the powerful tools of homological
algebra.

\section{Equations to First Order}
\setcounter{equation}{0}

In order to explain the cohomological approach, we shall first
write explicitly the conditions on the deformation
to first order in the coupling constant $g$.
We shall also write the triviality condition to
the same order.  The Noether identity
(\ref{FullNoether}) reads, to zeroth and first orders in $g$,
\begin{eqnarray}
O(g^0) &:& \; \frac{\delta I_0}{\delta \phi^i} R^{(0)i}_\alpha
= 0 \label{ZerothNoether}\\
O(g^1)  &:& \; \frac{\delta I_1}{\delta \phi^i} R^{(0)i}_\alpha +
\frac{\delta I_0}{\delta \phi^i} R^{(1)i}_\alpha = 0.
\label{FirstNoether}
\end{eqnarray}
The first condition (\ref{ZerothNoether}) is nothing
but the Noether identity (\ref{FreeNoether}) of the free theory
and is fulfilled by assumption.  The second condition
states that the gauge variation (for the undeformed
gauge symmetries) of the first-order deformation should
vanish when the (free) equations of motion hold,
\begin{equation}
\delta_\epsilon I_1 \approx 0.
\label{observables}
\end{equation}
Solutions of (\ref{observables}) are called ``observables".
First order deformations are therefore observables.
The relevant observables are of course ``integrated
observables", i.e., spacetime integrals of  local functions.

Similarly, it follows from (\ref{trivialdeformation})
that a first order-deformation is trivial if and only if it vanishes
on-shell (for the free theory),
\begin{equation}
\hbox{$I_1$ is trivial if and only if } \; I_1 \approx 0.
\label{FirstTrivial}
\end{equation}

We thus see that to first order in the deformation parameter $g$,
the problem of finding the consistent deformations is equivalent 
to that of finding the (integrated) observables of the undeformed
theory, with the understanding
that two observables that coincide on-shell are equivalent
-- as it is usually implied in the definition of
``observables".

It should be noted that the equations (\ref{observables})
and (\ref{FirstNoether}) are equivalent.  This is because
any function that is zero on-shell can be written as
a combination of the equations of motion (we assume the
necessary regularity conditions that guarantee this).  Thus,
to first order in $g$, it is only necessary to find the deformation
$I_1$ of the action.  The deformation of the gauge symmetry
follows from (\ref{observables}), (\ref{FirstNoether}) and
can be read off from the
coefficients of the equations of motion in the
variation of $I_1$.
Similarly, (\ref{FirstTrivial}) imply (\ref{trivialdeformation})
to first order in $g$.

Now, it is well known that the observables of a gauge theory
can be described cohomologically in terms of the BRST differential.
To correctly implement the equivalence relation
implied by the equations of motion, it is necessary to
include the antifields (Zinn-Justin ``sources 
for the BRST variations" \cite{ZJ}) and to
work within the antifield formalism developed by Batalin and
Vilkovisky \cite{BV} along lines initiated  
in \cite{Kallosh,deWit}.  

More precisely, one of the main 
points of the cohomological approach to the
construction of consistent interactions is that
there exists a differential $s$, the so-called
``BRST differential", whose cohomology in ``ghost
number zero" is precisely given by the space
of observables, on-shell vanishing observables being
BRST exact.  Thus, the non-trivial first-order deformations
are described by the group $H^0(s)$ (in the space of
local functionals).

The description of the observables of a theory involves two ingredients:
one is the gauge-invariance condition; the other is the
fact that an on-shell vanishing function should be identified
with zero.
As shown in \cite{FH,HProcSuppl,HT}, there corresponds a
separate differential for each of these ingredients.  The first
is the longitudinal differential $\gamma$ along the gauge
orbits, which implements the gauge invariance
condition.  The second is the Koszul-Tate differential $\delta$
associated with the stationary surface, which 
takes the equations of motion
into account.  The BRST differential combines these
two differentials into a single object (in the simplest
cases, it is just the sum).  We refer the reader
to \cite{FH,HProcSuppl,HT} for the details.  Locality is taken
care of in \cite{MHcmp}.  It is precisely because the BRST differential
contains the Koszul-Tate piece  -- something that does not appear
to be always properly appreciated -- that the above on-shell
relations are enforced when going to the BRST cohomology
\cite{FH,HProcSuppl,HT}.  The associated antifields, initially introduced
in order to keep an hand on the renormalization of the
BRST symmetry, have also the extremely important homological
interpretation of being the generators of the ``stationary"
Koszul-Tate
complex.

We shall give here only the form of $s$ for the specific
models given above.  In both these models, $s$ is simply
given by the sum of $\delta$ and $\gamma$, because the
gauge symmetries close off-shell.

In the case of reducible theories, one must impose in
addition to (\ref{FirstNoether}) the condition that the
gauge symmetries should remain reducible in the
deformation, i.e., that
(\ref{FullReducibility}) holds in the deformed theory to
order $g$.  This is, however, a consequence of (\ref{FullReducibility})
at this order, so that the requirement that $I_1$ be an observable
of the free theory is the sole independent requirement at order
$g$.  Indeed, if one contracts (\ref{FirstNoether})
with $Z^{(0)\alpha}_A$, uses the reducibility identity
at order zero, and recalls that the gauge transformations
of the free theory are assumed to form a complete set,
one easily finds that there exist functions $Z^{(1)\alpha}_A$
and $\mu_{A}^{(1)ij}$ such that (\ref{FullReducibility})
holds up to order $g$ (included) \cite{HT}.  Thus, the only condition at
order $g$ is (\ref{observables}), so that the non trivial
deformations are parametrized by the cohomogy group $H^0(s )$
(in the space of local functionals) also in the reducible case.

\section{BRST Differential}
\setcounter{equation}{0}

We shall from now on give up the
condensed notations where a summation over repeated
indices involved also an integral.  Spacetime integrals will
always be explicitly written and the objects that we shall
manipulate (``local functions")
will be ordinary functions of the variables
(original fields, ghosts,
antifields) and their derivatives up to some finite order,
without $\delta$-function or derivatives of it.  We shall also
deal with local spacetime forms, i.e., forms with coefficients
that are local functions.

The appropriate mathematical framework for dealing with local
forms is the one of jet bundles and it is straightforward to
formulate the general BRST cohomological construction in
this language \cite{jpaa}.  The reader may consult 
\cite{Vinogradov,Takens,Anderson,Schmid}
for information on jet bundles.

\subsection{BRST Differential For a Set of $U(1)$-Gauge Fields} 

By following the general prescriptions of the antifield formalism
\cite{BV,HProcSuppl,HT,Gomis}, one finds that the BRST differential
for a set of $U(1)$-gauge fields is given by
\begin{equation}
s = \delta + \gamma
\end{equation}
where the Koszul-Tate differential reads
\begin{equation}
\delta A^a_\mu = \delta C^a = 0, \; \delta A^{* \mu}_a = \partial_\nu
F^{\mu \nu}_a, \; \delta C^{*}_a = \partial_\mu A^{* \mu}_a
\end{equation}
while the exterior derivative along the gauge orbits is
\begin{equation}
\gamma A^a_\mu = \partial_\mu C^a, \; \gamma C^a = 0, \;
\gamma A^{* \mu}_a = 0, \; \gamma C^{*}_a =  0.
\end{equation}
In these relations, the $C^a$ are the ghosts, the
$A^{* \mu}_a$ are the antifields conjugate to
the vector potentials,
while the $C^{*}_a$ are the antifields conjugate to the ghosts.

The action of $\delta$ and $\gamma$ is extended to the
derivatives of the variables by demanding $\delta
\partial_\mu = \partial_\mu \delta$ and
$\gamma  \partial_\mu = \partial_\mu  \gamma $.  One then
extends the action of $\delta$ and $\gamma$  to
products of variables by using the Leibnitz rule
so that they are (anti)derivations.  It is easy
to check that
\begin{equation}
\delta ^2 = 0,\; \gamma^2 = 0, \; \delta \gamma + \gamma 
\delta = 0.
\end{equation}
Hence, $s$ is also a differential,
\begin{equation}
s^2 = 0
\end{equation}

The variables are conveniently assigned the following
gradings:

\vskip .3cm

\noindent
Antighost number:
\begin{equation}
antigh(A^a_\mu) = 0, \; 
antigh(C^a) = 0, \; antigh(A^{* \mu}_a) = 1, \;
antigh(C^{*}_a) = 2
\end{equation}

\vskip .2cm

\noindent
Pure ghost number:
\begin{equation}
puregh(A^a_\mu) = 0, \;
puregh(C^a) = 1, \; puregh(A^{* \mu}_a) = 0, \;
puregh(C^{*}_a) = 0.
\end{equation}
The (total) ghost number is the difference between the
pure ghost number and the antighost number.  Furthermore, the
$A^a_\mu$ and $C^{*}_a$ are even, while $C^a$ and
$A^{* \mu}_a$ are odd (anticommuting).

The Koszul-Tate differential $\delta$ decreases the
antighost number by one unit and does not modify
the pure ghost number.  The longitudinal derivative
$\gamma$ increases the pure ghost number by one
unit and does not modify the antighost number.
Accordingly, all three differentials $s$, $\delta$ 
and $\gamma$  increase the total ghost number by one unit.

\subsection{BRST Differential For a Set of Free Exterior $2$-Forms}

The rules for reducible systems yield the following BRST
differential $s = \delta + \gamma$ \cite{BV,HT,Gomis}
\begin{equation}
\delta B^A_{\mu \nu} = 0, \; \delta C^A_\mu = 0, \;
\delta \rho^A = 0, \; \delta B^{* \mu \nu}_A = 
\partial_\lambda H^{\mu \nu \lambda}_A, \;  \delta 
C^{* \mu}_A = \partial_\nu B^{* \mu \nu}_A, \;
\delta \rho^*_A = \partial_\mu C^{* \mu}_A
\end{equation}
and
\begin{equation}
\gamma B^A_{\mu \nu} = \partial_\mu C^A_\nu - \partial_\nu C^A_\mu, \;
\gamma C^A_\mu = \partial_\mu \rho^A, \; \gamma \rho^A = 0, \; 
\gamma B^{* \mu \nu}_A = 0, \; \gamma C^{* \mu}_A =  0, \;
\gamma \rho^*_A = 0.
\end{equation}
The $C^A_\mu$ are the ``ghosts" while the $\rho^A$ are the ``ghosts
of ghosts".  The $B^{* \mu \nu}_A$ are the
antifields conjugate to the $B$'s, the $C^{* \mu}_A$ are those
conjugate to the ghosts $C^A_\mu$, while the
$\rho^*_A$ are the antifields conjugate to the ghosts
of ghosts $\rho^A$.

The ghost number assignments are this time

\vskip .3cm

\noindent
Antighost number:
\begin{eqnarray}
& & antigh(B^A_{\mu \nu}) = 0, \; antigh(C^A_\mu) =0, \;
antigh(\rho^A) = 0,  \\
& & antigh(B^{* \mu \nu}_A) = 1, \;
antigh(C^{* \mu}_A) = 2, \; antigh(\rho^*_A) = 3
\end{eqnarray}

\vskip .2cm

\noindent
Pure ghost number:
\begin{eqnarray}
& & puregh(B^A_{\mu \nu}) = 0, \;  puregh(C^A_\mu) = 1, \;
puregh(\rho^A) = 2,  \\
& & puregh(B^{* \mu \nu}_A) = 0, \;
puregh(C^{* \mu}_A) = 0, \; puregh(\rho^*_A) = 0.
\end{eqnarray}
The total ghost number $gh$ is again the difference between the
pure ghost number and the antighost number.  One has $gh(s) = 1
= gh(\delta) = gh(\gamma)$.

By extending $\delta$ and $\gamma$ to derivatives and products as
above, one finds again that these are anticommuting differentials,
so that $s$ is also a differential.

\section{(Integrated) Observables}
\setcounter{equation}{0}

As we have pointed out, the set of observables is isomorphic
to the zeroth cohomology group of the BRST differential.
Since we are interested in integrated observables, we must
consider the cocycles and coboundaries of the BRST differential that
are also given by integrals of local densities (local
$n$-forms).  

If one works with the integrands $a$ of the integrated observables
$A = \int a$, which is more convenient, then one finds that
these should be in the so-called cohomology of $s$ modulo $d$ in
form degree $n$ (since $a$ is a local $n$-form).
That is, the cocycle condition $sA=0$ reads, in terms of $a$
\begin{equation}
sa + db = 0
\label{cocycle}
\end{equation}
for some $(n-1)$-form $b$, while a solution of (\ref{cocycle})
is trivial if and only if it can be written as
$a = sm +dl$, for some $m$, $l$.

The proof that the set of integrated local observables is isomorphic
with $H^0(s \vert d)$ is given in \cite{MHcmp,BBH1}.  A crucial step
in the proof is the acyclicity of the differential $\delta$ in the
space of local functionals with positive antighost
number and positive pure ghost number \cite{MHcmp}.

We shall not work out explicitly here the BRST cohomology $H^0(s \vert d)$
for the two models described above.  We shall merely report the results
and refer to the literature for the detailed proofs.
It is rather remarkable that the cohomological approach
gives the complete list of all first-order consistent
interactions for these (and other) models.  [That the vertices listed
below are consistent to first-order is rather obvious; that there are
no other vertices is perhaps not as straightforward and appears to be
a definite pay-off of the cohomological method].

\subsection{First-Order Consistent Interactions For a Set of
$U(1)$-Gauge Fields}

We shall classify the interactions according to whether
they deform non-trivially or not the gauge algebra, which is abelian
at order $g^0$.

Consistent interactions of a given gauge theory may be classified into
three categories: (i) those that do not modify the gauge transformations;
(ii) those that modify the gauge transformations without changing
their algebra; and (iii) those that modify both the gauge transformations
and their algebra.  For the first type, the gauge variation $\delta_\epsilon
I_1$ of the vertex $V$ vanishes (up to a surface term) off-shell and not just
on-shell.   For the second
and third types, $\delta_\epsilon I_1$ vanishes only on-shell,
\begin{equation}
\delta_\epsilon I_1 =  \int b^{a}_{\mu}
\frac{\delta I}{\delta A^{a}_{\mu}} d^n x
\label{VariationV}
\end{equation}
with $b^{a}_{\mu} \not= 0$.
The modification of the gauge transformations is given, to first order
in the coupling constant $g$, by
\begin{equation}
\delta^{NEW}_\epsilon A^{a}_{\mu} =
(d \epsilon^a)_{\mu} - g b^a_{\mu}
\label{GaugeTransf}
\end{equation}
since then, the gauge variation $\delta_\epsilon^{NEW} (I_0 + g I_1)$ vanishes
to order $g^2$.  If $b^a_\mu$ is gauge invariant, the second
variation $\delta_{\epsilon_1}
^{NEW} \delta_{\epsilon_2}^{NEW}  A^{a}$ is 
of order $g^2$ and the interaction
does not modify the gauge algebra to order $g$.

Interactions of each type exist for a set of free vector fields
$A^a_\mu$.  Let us start with the interactions that
truly deform (to first order in $g$) not only the gauge
transformations, but also their algebra.  In the
antifield language, these are the cohomological classes
of $H^0(s \vert d)$ for which all representatives necessarily
involve the antifields $C^*_a$ \cite{BH}.  As shown in 
\cite{BBH2,ijmp}, the only such interactions are given by
the familiar Yang-Mills cubic vertex proportional to
\begin{equation}
f_{abc} F^{ a\mu \nu} A^b_\mu A^c_\nu
\label{YMcubic}
\end{equation}
where the constants $f_{abc}$ are completely antisymmetric but otherwise
arbitrary at this stage.  The modification of the gauge
transformations induced to first order in $g$ is the familiar
transformation of a non-abelian gauge connection
\begin{equation}
\delta_\epsilon A^a_\mu = \partial_\mu
\epsilon^a + g f^a_{ \;bc} A^b_\mu \epsilon_c.
\end{equation}
The new gauge transformations involve explicitly
the vector potentials and no longer commute.  Their
algebra is the well-known Yang-Mills algebra.
Any gauge algebra deforming interaction
differs from (\ref{YMcubic}) by terms that do not
deform the algebra.  It is in this sense that (\ref{YMcubic})
is the unique vertex deforming the algebra.

Let us now turn to the interactions that deform non trivially
the gauge transformations, but do not deform their algebra.
These are exhausted by the terms of the form
\begin{equation}
A^a_\mu j^\mu_a
\label{Ajvertex}
\end{equation}
where $j^\mu_a$ are gauge-invariant conserved ``currents"
(for the free theory), constructed out of the potentials and
their derivatives.
Since one has
\begin{equation}
\partial_\mu j^\mu_a = t^b_{a \mu} \frac{\delta I_0}{\delta A^b_\mu}
\label{conservation}
\end{equation}
for some gauge-invariant functions $t^b_{a \mu}$,
the gauge transformations become
\begin{equation}
\delta_\epsilon A^a_\mu = \partial_\mu \epsilon^a
+ g t^a_{b \mu} \epsilon^b
\end{equation}
and clearly remain abelian to order $g$ since
$ t^a_{b \mu}$ is gauge-invariant.  [One could have terms
involving the derivatives of the field equations in
(\ref{conservation}) with the same conclusions].

As a free theory, the system of abelian gauge fields possesses
an infinite number of conserved currents.  Thus, there are
in general an infinite number of first-order-consistent
interactions that do not deform the algebra but do deform
the gauge transformations.  However, in $4$ dimensions, according
to a result due to Torre, there is no conserved current $j^\mu_a$ that
transforms as a Lorentz vector and does not involve explicitly
the coordinates \cite{Torre}.  Thus, there is no Lorentz-invariant
interaction of the type (\ref{Ajvertex}) in $4$ dimensions.
It seems plausible to extrapolate this result to all spacetime
dimensions, except $3$, where the gauge-invariant currents
\begin{equation}
g_{abc}\;  \epsilon^{\mu \beta \gamma} \; ^*\! F^a_\beta \; ^*\! F^b_\gamma
\end{equation}
are conserved and transform as Lorentz-vectors.
The corresponding interaction vertex
\begin{equation}
g_{abc}\;  \epsilon^{\mu \beta \gamma} \; A^a_\mu
\; ^*\! F^b_\beta \; ^*\! F^c_\gamma
\end{equation}
is known as the Freedman-Townsend vertex \cite{FT}.  Here,
$^*\! F^b_\beta$ is the one-form dual to the two-form
$F^b_{\mu \nu}$. 

Finally, the interactions that do not deform the gauge transformations
are given by the functions of the curvatures $F^a_{\mu \nu} \equiv
\partial_\mu A ^a_\nu - \partial_\nu A ^a_\mu$ and their derivatives,
as well as by the Chern-Simons terms 
\begin{equation}
g_{a_1 \dots a_k} A^{a_1} \wedge F^{a_2} \wedge
\cdots \wedge F^{a_k}
\end{equation}
in odd spacetime dimensions $n = 2k -1$ \cite{DJT}. Here,
$g_{a_1 \dots a_k}$ is completely symmetric.

The most general Lorentz-invariant, first-order consistent, interaction
vertex for a set of free abelian vector fields
is a linear combination of the Yang-Mills cubic vertex and
of the Lorentz-invariant, gauge-invariant functions (plus the
Freedman-Townsend vertex in $3$ dimensions and the Chern-Simons
terms in odd dimensions).  Higher-order
consistency will be discussed below.

\subsection{First-Order Consistent Interactions For a Set of
Free exterior $2$-Forms}

Consistent interactions for a set of exterior 
$2$-forms can also be classified according 
to whether they modify the gauge transformations
or the gauge algebra.   However, in the $2$-form case, the situation is
much simpler than for $1$-forms.  Indeed , there is {\em no}
interaction that truly deforms the gauge algebra.  Furthermore,
there is even no interaction that deforms non trivially the gauge 
transformations, except in $4$ dimensions, where the 
Freedman-Townsend vertex is the only possibility \cite{MHplb},
\begin{equation}
f_{ABC}\;  \epsilon^{\mu \nu \beta \gamma} \; B^A_{\mu \nu}
\; ^*\! H^B_\beta \; ^*\! H^C_\gamma.
\label{Freed}
\end{equation}
Here, $^*\! H^B_\beta$ is the one-form dual to the three-form
$H^B_{\mu \nu \lambda}$.  The Freedman-Townsend vertex deforms the gauge 
transformations as follows
\begin{equation}
\delta_\epsilon B^A_{\mu \nu} = D_\mu \epsilon^A_\nu
- D_\nu \epsilon^A_\mu, 
\label{newgaugeFT}
\end{equation}
where
\begin{equation}
D_\mu \epsilon^A_\nu = \partial_\mu \epsilon^A_\nu
+ g f^A_{\; BC} \, ^*\! H^B_\mu \epsilon^C_\nu.
\end{equation}

The new gauge transformations (\ref{newgaugeFT})
are clearly still reducible, but
only weakly so,
\begin{equation}
\delta_\epsilon B^A_{\mu \nu} \approx 0
\end{equation}
for
\begin{equation}
\epsilon^A_\mu = D_\mu \Lambda^A .
\end{equation}
The deformed coefficient $\mu^{ij}_A$ occuring
in the reducibility identities are non zero.

Thus, except in four dimensions, the only available
first-order consistent interactions do not deform the
gauge transformations and are given by gauge-invariant
functions of the field strength components and their
derivatives, as well as by Chern-Simons terms
\begin{equation}
f_{A_1 \dots A_k} B^{A_1} \wedge H^{A_2} \wedge
\cdots \wedge H^{A_k}
\end{equation}
in $2$
mod $3$ dimensions.  Here, $f_{A_1 \dots A_k}$ is completely
antisymmetric.  In particular, there is no
analog of the Yang-Mills cubic vertex coupling for
$2$-forms.  One does not need to invoke
Lorentz invariance to reach this result.  It is a direct
consequence of the cohomological analysis alone.
The gauge symmetries of exterior forms of degree $> 1$ 
are extremely rigid.  This was anticipated in \cite{Teitelboim}

\section{Consistent Deformations : Higher Orders}
\setcounter{equation}{0}

The BRST cohomology plays a central role to first order
in the deformation parameter because it gives, at ghost number
zero, the first order terms of the consistent deformations.
It also plays a central role at higher order in $g$, because
it is the group $H^1(s \vert d)$ that controls the
obstructions to the existence of the higher order deformation
terms \cite{BH}. 

The most expedient way to analyse this problem is through the master
equation \cite{ZJ,BV}.  We refer the reader to \cite{BH} for
the details and sketch here only the main idea.  

The space of fields, ghosts and antifields is naturally equipped by
an ``antibracket" structure, in which the antifields are
conjugate to the corresponding fields or ghosts.  This
antibracket structure induces an antibracket in the
local BRST cohomological classes $H^k(s \vert d)$.  It
is such that the antibracket of one element of $H^k(s \vert d)$
with one element of $H^j(s \vert d)$ is an element
of $H^{k + j +1}(s \vert d)$.  In particular, the
antibracket of one element of  $H^0(s \vert d)$
with one element of $H^0(s \vert d)$ is an element
of $H^1(s \vert d)$.  Now, a first-order consistent deformation
-- that is, an element of $H^0(s \vert d)$ -- is non obstructed to
second order if and only if its antibracket with itself
is BRST-exact, i.e. is the zeroth element of $H^1(s \vert d)$. 
This follows from a direct analysis of the
deformed solutions of the 
master equation,
\begin{equation}
(S,S) = 0
\label{master}
\end{equation}
with 
\begin{equation}
S = S^{(0)} + g S^{(1)} + g^2 S^{(2)} + O(g^3).
\end{equation}
We recall that $S$ contains all the information about the
action, gauge symmetries, gauge algebra ... of the
theory \cite{HProcSuppl,HT}.  From
(\ref{master}), one gets at order $g^2$ 
the condition
\begin{equation}
(S^{(1)}, S^{(1)}) + 2 (S^{(0)}, S^{(2)}) = 0.
\end{equation}
Since $S^{(0)}$ generates the BRST symmetry of the undeformed
theory through the antibracket, one sees that $S^{(2)}$ exists
if and only if the cocycle $(S^{(1)}, S^{(1)})$ is trivial in
$H^1(s \vert d)$.  If $(S^{(1)}, S^{(1)})$ is not BRST exact,
there is no $S^{(2)}$ and the deformation gets obstructed at
order $g^2$.  It is thus the cohomological group $H^1(s \vert d)$
that controls the obstructions to the existence of the second
order terms, as it was announced above.

It turns out that the obstructions to the existence of the higher order
terms besides $S^{(2)}$ can also be expressed in terms
of $H^1(s \vert d)$.  Thus, the problem of consistent deformations is
entirely governed by the BRST
cohomological groups $H^0(s \vert d)$ (for the first-order
deformations) and $H^1(s \vert d)$ (fopr the obstructions to
continuing the deformation to higher orders).  In particular,
if $H^1(s \vert d)$ vanishes, no first-order deformation can be 
obstructed.

The analysis of the higher-order consistency
conditions for the Yang-Mills deformation of the abelian 
$U(1)^N$ gauge theory
leads to the conclusion that the $f_{abc}$ should fulfill
the Jacobi identity and are thus the structure constants of
a Lie algebra.  If they do not fulfill the
Jacobi identity, the first order deformation (\ref{YMcubic}) 
is obstructed
at second order because the bracket of the corresponding
BRST cohomological class with itself is then
a non trivial element of $H^1(s \vert d)$ \cite{ijmp}.
If the Jacobi identity is fulfilled, the bracket is
zero (in cohomology) and $S^{(2)}$ is the familiar
Yang-Mills quartic contact term.

If one adds also gauge invariant terms or Chern-Simons
terms to the action, these get modified at higher order
to the corresponding non-abelian gauge-invariant
or Chern-Simons terms. [There is, for the Chern-Simons terms, a
second-order condition that the coefficients $g_{a_1 \dots a_k}$
should be invariant tensors of the deformed Lie algebra].

We have seen that in three dimensions, the
Freedman-Townsend vertex is another
candidate deformation.  By itself (or in
the presence of gauge invariant /Chern-Simons terms),
this vertex is not obstructed at higher order provided
the g's defining it fullfill also the Jacobi identity.
The higher order deformations exist and the full
interacting action is non polynomial (in this
second order form) \cite{FT}.

If one considers simultaneously, in three dimensions, the
Yang-Mills vertex and the Freedman-Townsend vertex, one
finds further conditions on the $f$'s and the $g$'s which
have been examined in \cite{Anco}.

For a set of free abelian $2$-forms, one finds that
the Freedman-Townsend interaction, which exists only in four
dimensions, is not obstructed only if the $f$'s fulfill
the Jacobi identity, as in three dimensions \cite{MHplb}.  The
resulting theory is non-polynomial.  In other spacetime dimensions
one finds of course that the gauge-invariant interactions,
which do not deform the gauge symmetries, are not
obstructed at higher order since the sum $I_0 + g I_1$
is in that case fully consistent to all orders.

\section{Other Models}
\setcounter{equation}{0}

The cohomological approach has been used in the analysis of other
models.  Although we have often called above the undeformed theory
the ``free theory", nowhere was it used that its Lagrangian 
was quadratic.  Accordingly, the formalism applies equally well to the 
investigation of the rigidity of an already interacting theory.
An important application 
is perturbative renormalization theory 
where the deformation
parameter is $\hbar$, which can be couched in
the antifield language \cite{VT,GW,W}.  

Various models have been studied.  The local BRST cohomology of
Yang-Mills theory has been worked out in \cite{BHPRL,BBH2}, following the
antifield-independent work of
\cite{DuboisViolette1,CohoH,Hann,DuboisViolette}.
The cohomological approach has enabled one to rigorously establish
previously unproved conjectures on the renormalization
of gauge invariant operators \cite{Kluberg,Joglekar}.

The rigidity of Einstein theory
of gravity against deformations that would modify
its gauge symmetries has been established in \cite{BBH3} (see also
\cite{Wald}).  The study applies also to
higher derivative models, whose quantum properties
have been investigated in
\cite{Stelle}.

Similarly
$D=4, N=1$ supergravity has been analysed in \cite{Brandt}
and the cohomological investigation of $D=11$-supergravity 
\cite{001} has been
started in \cite{Bautier}.  In that model, the
impossibility of introducing a cosmological constant
along conventional lines \cite{Sagnotti,NTvN} has
been explicitly proved and related to the fact that the
mass term for the spin $3/2$-field, which must accompany the
cosmological term, is not an observable.

Other recent applications include couplings of
$p$-forms \cite{HK,HKS,BrandtDragon,Bresil} and two-dimensional
models \cite{Brandt1,Brandt2}.  In particular, the
cohomological approach enables one to give the exhaustive
list of all the consistent interactions among
exterior form gauge fields of different degrees $\geq 2$.
It would be of interest to 
investigate the long-standing problem of interactions for massless
higher spin particles 
(e.g. spin 3) \cite{Berends1,Berends2,dWF,Bengtsson,Deser}
along the cohomological lines.  It is planned to pursue this
question in the future \cite{collab}.

\section{Conclusions}
\setcounter{equation}{0}

In this review, we have described the cohomological
approach to the problem of consistently deforming a gauge
invariant action.  We have indicated that the consistent
deformations are controlled by the local cohomological
groups $H^0(s\vert d)$ and $H^1(s\vert d)$ of the BRST differential.
The first group gives the first-order deformations.
The second contains the obstructions to higher-order
deformations.  We have also given references to articles where
explicit models are completely analysed along the cohomological lines.

\section*{Acknowledgements}
The author is grateful to Glenn Barnich, Friedemann Brandt,
Stanley Deser, 
Michel Dubois-Violette, Bernard Knaepen, Jim Stasheff, Michel
Talon and Claude Viallet
for useful discussions on matters related to this paper.

\end{document}